%Paper: solv-int/9411005
%From: NERNEY@ssl.msfc.nasa.gov
%Date: Thu, 17 Nov 1994 16:33:09 -0600 (CST)

\leftskip=0pt \parskip=0pt \parindent=0pt
\tolerance 10000
% \font\eightrm=cmr8
% \font\tenrm=cmr10

\font\twelverm=cmr10 scaled 1200
\font\fourteenbf=cmbx10 scaled 1440
\nopagenumbers
\headline={\tenrm\hfil\folio}
%\vskip .25in
\baselineskip=18pt
%\baselineskip = 2\baselineskip
%\voffset=2\baselineskip
\def\p#1 #2 {{{\partial #1}\over {\partial #2}}}
\def\ov#1 #2 {{{#1}\over {#2}}}
\vbox {
\centerline {\fourteenbf LIMITS TO EXTENSIONS OF BURGERS EQUATION}
\vskip .2in

\centerline {\twelverm Steven Nerney$^{1}$, Edward J. Schmahl$^{2}$
, and Z. E. Musielak$^{3}$}
\bigskip
$^1$ National Research Council Associate, NASA-Marshall Space Flight
Center, Alabama 35812.

$^2$ Astronomy Department, University of Maryland, College Park,
MD 20742.

$^3$ Department of Mechanical and Aerospace Engineering,
and Center for Space Plasma and Aeronomic Research, University
of Alabama at Huntsville, Huntsville, Alabama 35899.\hfill
\bigskip
Accepted for publication in the Quarterly of Applied Mathematics\hfill
May 31, 1994.\hfill
\bigskip
\vskip .4in
%{\parindent =12pt \narrower\narrower\narrower}
\leftskip = 1in \rightskip =1in
{\bf Abstract.}
The vector Burgers equation is extended to include pressure
gradients and gravity. It is shown that within the framework
of the Cole-Hopf transformation there are no physical solutions
to this problem. This result is important because it
clearly demonstrates that any extension of Burgers equation
to more interesting physical situations is strongly limited.
}
\eject
%\vskip .4in
\baselineskip = 2\baselineskip
{\bf 1. Introduction}
\vskip .1in
We solved Burgers equation in three dimensions for arbitrary
orthogonal curvilinear coordinate systems [1]; hereafter, NSM
(also see [2]). The analytical solutions suffered
from the well-known deficiencies of Burgers equation, which
does not account for gravity and pressure gradients, nor does
it conserve mass. In this paper, we have attempted to extend
Burgers equation by including gravity and either isothermal or
adiabatic pressure gradients. The basic idea of our approach
is to split the momentum equation into the vector Burgers
equation and the equation which contains all remaining terms,
and to solve these two equations simultaneously. We show that
the latter can be done only when a special auxillary condition
is satisfied. Because the solutions to the vector Burgers
equation are well-known [1], the main problem is to find
physically acceptable solutions to the auxillary condition
that also satisfy the continuity equation. We have attempted
to generate steady-state, viscous solutions by constructing
them from the solution of Laplace's equation in such a way
as to satisfy the continuity equation for compressible flow.
Our apparent success is vitiated when we find that there are
no physically acceptable solutions to the 2-D
viscous case that can be derived from the potential flow
problem and produce physical pressure gradients. This
is an important result because it shows that the obvious extension
of the vector Burgers equation to more interesting physical
cases is strongly limited. Still, the solutions are of interest in
the same sense that the analytic solutions of the vector Burgers
equation are interesting.
\smallskip
We begin our presentation by summarizing the assumptions and
solutions of the vector Burgers equation presented in paper [1]
and then describe in detail our new results.
\vskip .1in
{\bf 2. Basic Equations}
\vskip .1in
The vector Burgers equation can be written in the followin form:
$$ \p  {\vec u} t  + \vec u\cdot \nabla \vec u = \nu \nabla ^2\vec u.
\eqno (1)$$
We proved that the solution of this equation can be expressed as a
function of the solution of the diffusion equation:
$$ \p {\Theta } t  = \nu \nabla ^2\Theta ,\eqno (2)$$
 namely,
$$ \vec u = -\ov {2\nu } {\Theta } \thinspace \nabla \Theta
.\eqno (3)$$
This result is correct for arbitrary curvilinear coordinate systems
and is a subset of the more general solutions to the tensor Burgers
equation [2]. We would like to be able to extend our previous solutions
to include pressure gradients and gravity, as well as satisfying the
continuity equation. Then we must solve the momentum equation,
$$ \p  {\vec u} t  + \vec u\cdot \nabla \vec u = -\ov {\nabla }P
{\rho }  + \nabla \ov {GM} r + \nu \nabla ^2\vec u + \vec f
\eqno (4)$$
where $\vec f$ is a general conservative force per unit mass so that
$$ \vec f = \nabla \Phi _f. \eqno (5)$$
and $\rho ,P, G, M$ are density, pressure, the gravitational
constant, and the mass of the object exerting the gravitational pull,
respectively.
The continuity equation is written as
$$ \p {\rho } t + \nabla \cdot (\rho \vec u) = 0. \eqno (6)$$

In order to evaluate the vector Laplacian for arbitrary orthogonal
coordinate systems, we must use
$$ \nabla ^2\vec u = \nabla (\nabla \cdot \vec u) - \nabla \times
(\nabla \times \vec u),\eqno (7)$$
and for the inertial term
$$ \vec u\cdot \nabla \vec u = \nabla \ov  {u^2} 2  - \vec u\times
(\nabla \times \vec u).\eqno (8)$$
Then eq. (4) greatly simplifies as long as
$$\vec u\times (\nabla \times \vec u) = \nu \nabla \times (\nabla
\times \vec u).\eqno (9a)$$
This appears to be an equation that represents quite general flows,
but the only solutions are irrotational. This can be seen by taking
the divergence of both sides of eq. (9a) and using
$$\nabla \cdot (\vec A\times \vec B) = \vec B\cdot (\nabla \times
\vec A) - \vec A\cdot (\nabla \times \vec B).\eqno (9b)$$
Therefore we require
$$ \nabla \times \vec u = 0,\eqno (10)$$
and
$$ \vec u = \nabla \Phi .\eqno (11a)$$
\medskip
We also use
$$ \ov {\nabla P} {\rho } =\nabla \left(\int \ov {dP} {\rho } \right)
\eqno (11b)$$
so that all terms in eq.(4) are the gradient of a scalar. The integral
of the pressure term can be done whenever $P=P(\rho )$, but
we will solve eq. (4) in the limit of an isothermal pressure gradient
with the
ideal gas law ($P=a^2\rho $, where a is the isothermal sound speed).
Then the pressure gradient term may be written as
$$\ov {\nabla P} {\rho } = a^2\nabla \ln \rho .\eqno (12)$$

Now, using eqs. (9a), (11a), and (12), eq. (4) simplifies to
$$ \nabla \left [\p {\Phi } t  + {{(\nabla \Phi )^2}\over {2}}
+ a^2\ln \rho  - \ov {GM} r - \nu \nabla ^2\Phi - \Phi _f\right ] =
0.\eqno (13)$$
This implies that the integral of eq. (13) is only a function of time,
so that
$$ \p {\Phi } t  + {{(\nabla \Phi )^2}\over {2}}
+ a^2\ln \rho -\ov {GM} r - \nu \nabla ^2\Phi - \Phi_f = E(t).
\eqno (14)$$
\medskip
The following vector identity is useful and leads to an important
form for $\nabla ^2\Phi $:
$$ \nabla \alpha  = \p  {\alpha } {\Theta } \nabla \Theta ,\eqno (15a)$$
which is true in any orthogonal coordinate system.
Taking the divergence of both sides with $\alpha =\Phi $ and using eq.
(15a) again with $\alpha =\ov {d\Phi } {d\Theta }  $ yields
$$ \nabla ^2\Phi  = \ov {d^2\Phi } {d\Theta ^2}  \thinspace
(\nabla \Theta )^2 + \ov {d\Phi } {d\Theta }  \thinspace \nabla ^2
\Theta ,\eqno (15b)$$
which, again,  is valid for any orthogonal curvilinear coordinate
system. Substituting eq. (15b) into (14)
$$ \p {\Phi } t  + {{(\nabla \Phi )^2}\over {2}}
+ a^2\ln \rho -\ov {GM} r  - \Phi_f -\nu \left(\ov {d^2\Phi }
{d\Theta ^2}
 (\nabla \Theta )^2 +\ov {d\Phi } {d\Theta }  \nabla ^2\Theta
\right) = E(t).\eqno (16a)$$
Now eq. (2) guarantees that the first and last terms on the
left-hand-side will cancel so that
$$ (\nabla \Theta )^2 \left[\ov 1 2 \left(\ov {d\Phi } {d\Theta }
\right)^2 - \nu \ov {d^2\Phi } {d\Theta ^2} \right]+a^2\ln \rho
-\ov {GM} r -\Phi_f=E(t). \eqno (16b)$$
An important aspect of our work in NSM was that the remaining terms
in the equation similar to eq. (16b) were proportional to $(\nabla
\Theta )^2$ so that this factor cancelled out, leading to an
integrable equation for $\p {\Phi } {\Theta } $.
It is clear that the
same procedure will work here provided that the following auxillary
condition is satisfied:
$$ a^2\ln \rho  -\ov {GM} r -\Phi _f-E(t) = \pm C^2(\nabla
\Theta )^2. \eqno (17a)$$
It is easy to show that C must be found from the solution of
$$C^2(\Theta )= \nu \ov {d^2\Phi } {d\Theta ^2}
-\ov 1 2 \left(\ov {d\Phi } {d\Theta } \right)^2, \eqno (17b)$$
but this is not particularly helpful.
We choose to examine possible solutions for the case of constant C
partly because C=0
corresponds to the vector Burgers equation solution discussed by NSM,
and we are attempting to find physical extensions of the Burgers
equation solutions. We are also physically motivated by noticing
that the particular choice of $C^2=-1/2$ in eq. (17a)
reduces the auxillary condition to the Bernoulli
equation which must be solved for the pressure distribution implied
by a potential flow model (the steady limit of the diffusion equation).
 The difference is that the auxillary condition is for compressible
flow. Additionally, as we will also show, the case of constant C
generalizes known velocity
laws derived from the solution of Burgers equation. It is for
these reasons
that we examine the solutions of the irrotational momentum equation
subject to the constraint of eq. (17a).
\medskip

The differential equation that results from substituting
eq. (17a) into eq. (16b) (the analog of eq. (7a) in NSM) is
$$\left (\ov {d\Phi } {d\Theta } \right )^2 = 2\nu \ov {d^2\Phi }
{d\Theta ^2}  \mp 2C^2.\eqno (18)$$
We solve for
$\p {\Phi } {\Theta } $ and use
$$ \vec u=\nabla \Phi = \ov {d\Phi } {d\Theta } \nabla \Theta ,
\eqno (19a)$$
to find
$$ \ov {d\Phi } {d\Theta }  = \sqrt 2C\tan \left(\ov C {\sqrt2\nu }
\Theta +C_1\right),\eqno (19b)$$
$$\vec u =  \sqrt 2C \tan \left(\ov C {\sqrt2\nu } \Theta +
C_1\right)\nabla \Theta ,\eqno (19c)$$
where $C_1$ is a constant of integration.  This solution corresponds
to the minus sign in eq. (18) while
$$\vec u = \sqrt 2C \tanh \left(\ov C {\sqrt2\nu } \Theta + C_1\right)
\nabla \Theta ,\eqno (19d)$$
corresponds to the positive sign. The first solution seems
nonphysical because of the singularities, although solutions could
be generated by restricting the argument between 0 and  $\pi /4$.
But, as we will show shortly, the existence of any physical solutions
require choosing the hyperbolic tangent solution. It is also known
[3,5] that the limit of eq. (19d) with $\nabla \Theta $=constant
is the steady solution to the 1-D cartesian
Burgers equation for an initial step function profile with a sharp
but finite discontinuity in u at x=a which rounds off asymptotically
in time to a steady-state shock layer profile of thickness $\nu /u_0$:
$$u=-u_0\tanh \left[\ov {u_0(x-a)} {2\nu } \right].\eqno (19e)$$
The tangent form is also known to be a steady-state solution of the
cartesian Burgers equation [5].
\medskip
The obtained solution, eq. (19d), leads to a mechanical energy
equation which is found by taking the dot product of eq. (19d)
with $\vec u$ and then eliminating $(\nabla \Theta )^2$ in eq.
(17a). This gives
$$ \ov {u^2} 2 \coth ^2\left(\ov C {\sqrt2\nu } \Theta +C_1\right)- \ov
{GM} r +a^2\ln \rho -\Phi _f=E(t).\eqno (20)$$
Note that although this equation is similar to Bernoulli's
equation, eq. (20) is for nonsteady, viscous flow. The
hyperbolic tangent solution can be seen as the appropriate
form because the internal energy/gram will decrease if the
speed increases. The form of the mechanical energy equation
is strongly reminiscent of nonlinear decay models [1]. A
singular source of mass at the origin requires solutions for
$\coth \Theta $ that begin at finite radius close to the source,
but then the motion decays so that the kinetic energy/mass
approaches zero as t$\rightarrow \infty $ when E(t)$\rightarrow 0$.
\medskip
We must now impose mass conservation on this model, and, to this
end, we find a useful form for the continuity equation. Using eq.
(11a) and further eliminating $\nabla \cdot \vec u$ by using eqs.
(2) and (15b) yields:
$$ \left(\p  {} t  + \vec u\cdot \nabla \right )\ln \rho
 + \ov {d^2\Phi } {d\Theta ^2}  (\nabla \Theta )^2 + \ov {1} {\nu }
\p {\Phi } {t} =0.\eqno (21)$$
\medskip
To evaluate the continuity equation, we use streamline coordinates
$$\hat e_s=\ov {\vec u} {|\vec u|}. \eqno (22)$$

{\bf 3. Steady Solutions}
\medskip

We further limit physical space to steady solutions so that eq. (2)
reduces to Laplace's equation. Then eq. (19d) requires that
$\nabla \Theta $ is zero perpendicular to streamlines and $\nabla
\Theta $ is $\p {\Theta } {s} $.
The problem reduces to finding that subset of the solution of
Laplace's equation that satisfies the auxillary condition and conserves
 mass.
\medskip
Identifying $\nabla \Theta $ with a velocity derived from a
potential flow solution
$$\nabla \cdot \vec u_{pot} = 0. \eqno (23a)$$
$$\nabla \cdot (\rho \vec u) = 0. \eqno (23b)$$
so that
$$\rho \vec u=\rho_{pot}\vec u_{pot}+\nabla \times \vec A.\eqno (23c)$$
The following details show that $\nabla \times \vec A$ is zero.
\medskip
The steady continuity equation is derived from eq. (21)
$$u\p { } {s} \ln \rho + \ov {d^2\Phi } {d\Theta ^2}  \left(\p
{\Theta } {s} \right )^2=0.\eqno (24)$$
Dividing by $\p {\Theta } {s} $ and using
$$u= \ov {d\Phi } {d\Theta }  \p {\Theta } {s} \eqno (25)$$
(for solutions with non-zero
velocities), eq. (24) integrates exactly with respect to
$\Theta $ to
$$ \rho \ov {d\Phi } {d\Theta }  =k_1 , \eqno (26)$$

where $k_1$ might be a function of spatial coordinates.
Substituting eq. (26) back into (19a)
$$\rho \vec u = k_1\nabla \Theta , \eqno (27a)$$
or, if the original solution of Laplace's equation is
formulated in terms of the velocity,
$$\rho \vec u = k_1\vec u_{pot} , \eqno (27b)$$
and $k_1$ is seen to be the original incompressible density
for the potential flow solution, $\rho _{pot}$, i.e., a strict
constant. An important aspect of this method is that the new
steady, compressible solutions are created from the old
potential flow (i.e., incompressible) solutions, provided that
the boundary condition on the compressible solution is formulated in
terms of the mass flux.
\medskip

Substituting the potential flow speed back into eq. (20)
yields another form for the steady mechanical energy equation:

$$ a^2\ln \rho  -\ov {GM} r -\Phi _f+ C^2u^2_{pot}=E_0.
\eqno (28a)$$

Setting $\Phi _f$ to zero and choosing
$$C^2=\ov 1 2 , \eqno (28b)$$
we get
$$ a^2\ln \rho  -\ov {GM} r + \ov {u^2_{pot}} {2} =E_{0}.
\eqno (29)$$
The auxillary condition is now seen to be the Bernoulli equation
that must be used to solve for the pressure distribution implied
by the potential velocity. However, we use eq. (19d) to solve for
the next higher-order solution for the velocity related to
this variable pressure distribution, but including viscosity.
\medskip
With this value for C, we obtain
$$\vec u=-\tanh \left(\ov {\Theta } {2\nu } +C_1\right)\enspace
\vec u_{pot} , \eqno (30a)$$
and
$$\rho =\rho _{pot}\coth \left(\ov {\Theta } {2\nu } +C_1\right),
\eqno (30b)$$
\medskip
The last part of this solution is to solve for the subset of
physical potential flow solutions. Setting gravity to zero and
substituting eq. (30b) into (28a), we find
$$u_{pot}^2=2E_{0}-2a^2\ln \coth \left(\ov {\Theta } {2\nu }
+C_1 \right) . \eqno (31a)$$
Replacing the $\int dP/\rho$ with the adiabatic term,
$(\gamma -1)P/\gamma \rho$, and substituting $\gamma =5/3$,
and then using eq. (26) gives the related equation for the
adiabatic case:
$$u_{pot}^2=2E_{0}-5b\coth ^{2/3}\left(\ov {\Theta } {2\nu }
+ C_1\right) , \eqno (31b)$$
where we have also used the adiabatic equation of state,
$$p=b\rho ^{\gamma }.\eqno (31c)$$
The potential flow solutions are limited to those whose speed is
constrained to obey eqs. (31a,b).

\medskip

We use complex variables to show that there are no physical
solutions that simultaneously satisfy both Laplace's equation
as well as the constraint on the speed, eqs. (31a,b). Writing
the complex potential, F(z), as
$$F(z)=\Theta (x,y)+i\Psi (x,y) , \eqno (32a)$$
with
$$ z=x+iy , \eqno (32b)$$
then
$$W(z)=\ov {dF} {dz} =v_x-iv_y , \eqno (32c)$$
where $\Psi $ is the stream function, and W(z) is the
complex velocity. The utility of this approach is obvious
because
$$WW^{\ast }=v_x^2+v_y^2 , \eqno (33)$$
where $W^{\ast }$ is the complex conjugate of W, which allows
the velocity components to be read from eqs. (31a) or (31b).
The method of complex variables requires that these velocity
components must be derived from an analytic potential function.
In the following, we demonstrate that this is not possible.

Consider the general form of eqs. (31a,b):
$$\left(\nabla \Theta \right)^2 = f(\Theta ).\eqno (35)$$
This equation implies that the gradient of $\left(\nabla
\Theta \right)^2$ must be parallel to $\nabla \Theta $, or that the
Jacobian of $\Theta $ and $\left(\nabla \Theta \right)^2$
must be zero. This is a necessary condition for solutions to eqs.
(31a,b) to exist.
The Jacobian simplifies (in cartesian coordinates) to
$$ (v_x^2-v_y^2)\p {v_x} y -2v_xv_y\p {v_x} x =0 , \eqno (36)$$
where we have also used the Cauchy-Riemann equations,
$$v_x=\p {\Phi } x =\p {\Psi } y ~, \eqno (37a)$$
$$v_y=\p {\Phi } y =-\p {\Psi } x ~. \eqno (37b)$$
But eq. (36) is the same as
$$Im\left(W^{\ast 2}(z)\ov {dW(z)} {dz} \right) = 0 .
\eqno (38a)$$
One solution is the trivial case of zero derivative. The other is
found by
defining G(z) to be the reciprocal of W(z), so that eq. (38a)
may be written as:
$$Im \left(\ov {dG/dz} {(G^{\ast } G)^2} \right)=0 .
\eqno (38b)$$
Because the denominator is real, we require, except at poles,
$$Im \left(\ov {dG(z)} {dz} \right)=0 . \eqno (38c)$$
But eqs.(37) then require
$$ \ov {dG(z)} {dz} =0 , \eqno (39)$$
so that
$$ W(z)=\ov c z , \eqno (40)$$
where c is a complex constant.
This gives the solution
$$ \Theta =a_1\ln r +a_2 , \eqno (41a)$$
together with the trivial solution
$$\Theta =a_3x+a_4y+a_5 , \eqno (41b)$$
where the $a_i$ are constants. As
eqs. (31a) and (31b) are not of this form, there are, then,
no two-dimensional solutions of the potential flow problem
that satisfy the physical constraint on the speed.
As these are
required to solve for physical viscous, compressible solutions
through eqs. (29) and (30a), there are no physical solutions
of this model. This is an important result because it shows
that the method of solving Burgers equation based on the
the Cole-Hopf transformation cannot be used to solve a
general (and physically more interesting) problem of a
flow with gravity and pressure gradients. To sum up,
substituting eq. (26) into eq. (16b) clearly indicates that
$(\nabla \Theta )^2$ is a function of $\Theta $. But, we have shown
that the only
steady solutions for $\nabla \Theta $ are those
generated from eqs. (41a,b). Our Cole-Hopf extension
of the solution of Burgers equation requires that
$F(\Theta )$ must be a function of $\coth (\Theta /2\nu +C_1)$,
so that there are no physical solutions.
\smallskip
{\bf 4. Conclusions}
\medskip
We have extended the solutions to the vector Burgers equation [1]
by a method that splits the momentum equation into two
pieces, each of which is satisfied simultaneously. This allows us
to solve a subset of irrotational flow problems
with pressure
gradients and gravity. The steady limit, when combined with mass
conservation,
reduces to specifying a subset of potential flow solutions. These
incompressible flows, in turn, allow the generation of viscous,
compressible solutions
that are related to the solutions of the vector Burgers
equation [1]. We then proved that the subset of two-dimensional
potential flow solutions
does not correspond to the requirements for physical flow derived
from real
pressure gradients. In this sense, the pressure gradients are derived
from pseudo-densities. It seems unlikely that physical models could
be generated by this method using either three
spatial dimensions or time-dependent solutions. Still, these results
are of interest in the same sense that the analytic solutions of the
vector Burgers equation are interesting.
\medskip {\bf Acknowledgments.}
\vskip .1in
ZEM acknowledges partial support of this work by the NSF under grant
no. ATM-9119580. EJS acknowledges partial support of this work
under grant no. NAG 2001 from NASA-Goddard Space Flight Center.
\vskip .1in
\centerline {References}
\vskip .1in
[1] S. Nerney, E. J. Schmahl, and Z. E. Musielak, {\it Analytic
solutions of the vector Burgers equation}, Quart. Appl. Math.,
in press, (1994).

[2] K. B. Wolf, L. Hlavaty, and S. Steinberg, {\it Non-linear
differential
equations as invariants under group action on coset bundles: Burgers
and Korteweg-de Vries equation families}, J. Math. Anal. and Applic.,
{\bf 114}, 340, (1986).

[3] J. D. Cole, {\it A quasi-linear parabolic equation in
aerodynamics}, Quart. Appl. Math., {\bf 9}, 225, (1951).

[4] E. Hopf, {\it The partial differential equation $u_t+uu_x=\mu
u_{xx}$}, Comm. Pure and Appl. Math., {\bf 3}, 201, (1950).

[5] E. R. Benton, {\it Solutions illustrating the decay of dissipation
layers in Burgers' nonlinear diffusion equation}, Phys. Fluids,
{\bf 10}, 2113, (1967).
\vfill
\eject
\bye